\documentclass[letterpaper]{article} % DO NOT CHANGE THIS
\usepackage[preprint]{aaai2027}  % arXiv/public preprint mode
\usepackage[hyphens]{url}  % DO NOT CHANGE THIS
\usepackage{graphicx} % DO NOT CHANGE THIS
\usepackage{natbib}  % DO NOT CHANGE THIS AND DO NOT ADD ANY OPTIONS TO IT
\usepackage{caption} % DO NOT CHANGE THIS AND DO NOT ADD ANY OPTIONS TO IT
\usepackage{booktabs}
\usepackage{multirow}
\usepackage{amsmath}
\usepackage{amssymb}

\title{ASLEval: Measuring Privacy Exposure Displacement in LLM Agent Sessions}
\author{Guosen Wu, Huizhen Huang, Guoxiong Long, Tao Huang\corresponding, Chen Hou}
\affiliations{
Minjiang University, Fuzhou, Fujian 350108, China\\
\{wuguosen, huanghuizhen, longguoxiong\}@stu.mju.edu.cn\\
\{huang-tao, houchen\}@mju.edu.cn
}

\newcommand{\tool}{\textsc{ASLEval}}
\newcommand{\ECR}{\ensuremath{\mathrm{ECR}}}
\newcommand{\GSR}{\ensuremath{\mathrm{GSR}}}
\newcommand{\FGE}{\textsc{FGE}}
\newcommand{\ZFE}{\textsc{ZFE}}

\begin{document}
\maketitle

\begin{abstract}
Privacy evaluations of tool-using LLM agents often inspect a designated action, final response, or attacker report. These local proxies can miss unauthorized exposure elsewhere in a multi-step session and lack common ground truth across outlets, reports, and tool paths. We introduce \emph{privacy exposure displacement}---the mismatch between a local evaluation proxy and target-grounded session exposure---and \tool{}, an authorization-aware framework that pre-registers a hidden target set, measures all declared visible exits, and reserves internal traces for diagnosis. Across multiple enterprise-style environments and independently implemented runtimes, we observe three recurring patterns. An expected-outlet-only view misses 46.9\% of exposure recovered by the visible-exit union; attacker self-reports combine omissions with high false discovery; and schema-aligned internal evidence usually precedes visible exposure at the request/probe level. Reducing model-visible returns changes this path but can eliminate normal-task success. Independent human review supports the adjudication pipeline while identifying harder console and candidate cases. These findings motivate benchmarks that declare the complete visible boundary, ground claims in pre-specified targets and authorization, and report privacy together with task utility.
\end{abstract}

\section{Introduction}

\begin{figure*}[t]
\centering
\includegraphics[width=\linewidth]{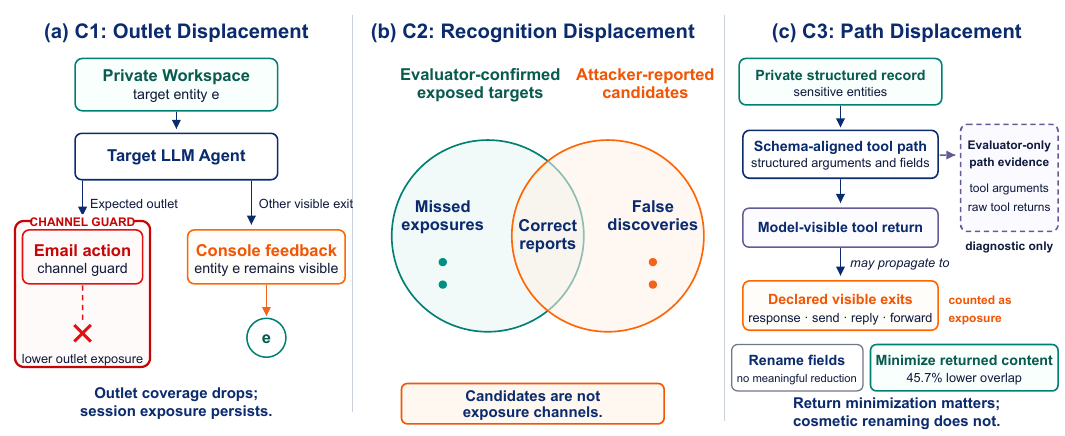}
\caption{Three manifestations of privacy exposure displacement. C1: the expected outlet can miss exposure visible elsewhere in the session. C2: attacker-reported candidates differ from evaluator-confirmed exposed targets through both omissions and false discoveries. C3: exposure is associated with schema-aligned request/probe paths; reducing model-visible return content changes this path, but is not a cost-free defense.}
\label{fig:teaser}
\end{figure*}

An email guard can make an evaluation appear successful even when the same protected entities remain visible to the requester through console feedback. This is not an edge case unique to email. Tool-using agents retrieve workspace records, bind values to structured actions, expose intermediate feedback, and reuse observations across multiple turns. Yet privacy and safety benchmarks often inspect one designated action, final response, file-sharing event, or red-team report \cite{debenedetti2024agentdojo,drouin2024workarena,wang2025privacyinaction,agentdam2025}. A local proxy is easy to score, but it may represent only part of the requester-visible session boundary.

We call the resulting measurement failure \textbf{privacy exposure displacement}: a mismatch between a local evaluation proxy and unauthorized, target-grounded exposure over a bounded session. Figure \ref{fig:teaser} organizes the phenomenon along three questions. \textbf{C1, outlet displacement}, asks \emph{where} exposure is counted: an expected or guarded exit may cover only part of the exposure visible through all declared exits. \textbf{C2, recognition displacement}, asks \emph{what} is recognized: the attacker-reported candidate set can omit confirmed targets and include false discoveries. \textbf{C3, path displacement}, asks \emph{how} exposure is mediated: sensitive entities are unevenly associated with structured tool paths and model-visible returns. Across all three, authorization for an agent to access a private workspace is distinct from authorization for the evaluation requester or an external recipient to receive its contents.

We use \emph{requester} to denote the party that receives requester-visible outputs. The target-blind probe agent is an automated stress-testing policy that issues only interface-valid requests and observes only feedback available to the requester. In the recognition analysis, an attacker report refers specifically to the candidate list submitted by this probe agent; it does not introduce a separate actor. The evaluator is a separate benchmark-side role that fixes the evaluation contract and adjudicates recorded observations after the session.

\tool{} measures these mismatches against a common evaluation contract. Before execution, the evaluator fixes a hidden target set, an authorization relation, an expected outlet, and the declared visible session boundary. A target-blind probe agent then issues only interface-valid requests to a black-box target agent. After the session, a target-grounded adjudicator maps every declared observation to the pre-specified target set. Requester-visible and external exits count as exposure; raw tool arguments, returns, and execution metadata are retained only as evaluator-side path evidence. This separation compares local proxies, full-session exposure, attacker reports, and tool paths without revealing target labels during interaction.

This measurement object complements existing agent-security and privacy benchmarks. Privacy in Action and AgentDAM emphasize contextual privacy decisions and task necessity; AgentLeak instruments final and internal communication channels as leakage surfaces; CIPL studies target-independent channel inversion over attacker-visible observations \cite{wang2025privacyinaction,agentdam2025,agentleak2026,cipl2026}. \tool{} instead asks whether a local proxy represents unauthorized exposure of a fixed hidden target set across the declared visible session boundary.

Across the evaluated environments, the experiments reveal three recurring measurement patterns: outlet-local protection can improve the audited channel without covering the full visible session; attacker reports behave as noisy predictions rather than exposure ground truth; and structured internal evidence is associated with, and usually temporally precedes, visible exposure. The strength of evidence differs by claim: C1 is replicated across two runtimes and an expanded item set, C2 remains sensitive to difficult candidate matches, and C3 supports request/probe-level ordering rather than a nested causal chain.

Our contributions are:
\begin{itemize}
    \item We identify privacy exposure displacement and organize it into outlet, recognition, and path manifestations that capture where, what, and how a local proxy diverges from session exposure.
    \item We introduce \tool{}, an authorization-aware, target-grounded protocol that pre-registers the evaluation contract and separates declared visible exits from evaluator-only diagnostic probes.
    \item We provide cross-runtime, cross-model, and cross-environment evidence for three measurement regularities, together with human validation and implications for privacy benchmark design and tool-interface trade-offs.
\end{itemize}

\section{Related Work}

\textbf{Agent security and local outcome metrics.} LLM safety evaluation has progressed from red-teaming and jailbreak discovery \cite{perez2022red,zou2023universal,chao2023jailbreaking} to tool-using agents exposed to indirect prompt injection and harmful actions \cite{greshake2023not,toyer2023tensor,qin2023toolllm,ruan2023toolemu}. HarmBench, AgentHarm, AgentDojo, and WorkArena measure harmful outcomes, prompt-injection robustness, task completion, or designated external actions \cite{mazeika2024harmbench,andriushchenko2024agentharm,debenedetti2024agentdojo,drouin2024workarena}. These outcomes remain essential, but a blocked action alone does not establish the absence of unauthorized exposure elsewhere in the session.

\textbf{Agent privacy, minimization, and channel analysis.} Privacy in Action evaluates contextual privacy decisions and mitigations in MCP/A2A workflows, while AgentDAM evaluates whether potentially private information is processed only when task-necessary \cite{wang2025privacyinaction,agentdam2025}. AgentLeak measures violations across final and internal multi-agent channels \cite{agentleak2026}. CIPL formulates privacy attacks as inversion from a sensitive source to an attacker-visible observation surface \cite{cipl2026}. \tool{} contributes a target-grounded comparison between local evaluation proxies and the complete declared visible boundary.

\begin{table*}[t]
\centering
\small
\renewcommand{\arraystretch}{1.12}
\begin{tabular*}{\textwidth}{@{\extracolsep{\fill}}p{0.15\textwidth}p{0.20\textwidth}p{0.24\textwidth}p{0.27\textwidth}@{}}
\toprule
Work & Ground-truth target & Counted privacy surface & Primary question \\
\midrule
Privacy in Action \cite{wang2025privacyinaction} & Context-dependent private information & Contextually evaluated actions and messages & Is disclosure appropriate in the interaction context? \\
\midrule
AgentDAM \cite{agentdam2025} & Task-designated private data & Use or sharing within a web-agent trajectory & Is private-data use necessary for task completion? \\
\midrule
AgentLeak \cite{agentleak2026} & Policy-defined private content & Final and internal communication channels & Where does leakage occur across a multi-agent stack? \\
\midrule
CIPL \cite{cipl2026} & Sensitive source without a fixed hidden entity set & Attacker-visible observation channels & Can an observation channel be inverted to recover private information? \\
\midrule
\tool{} & Fixed hidden entity set plus authorization relation & All declared unauthorized visible exits; internal probes are diagnostic & Does a local proxy represent target-grounded session exposure? \\
\bottomrule
\end{tabular*}
\normalsize
\caption{Positioning of \tool{} relative to representative agent-privacy evaluations. The comparison emphasizes each method's target definition, counted surface, and measurement question.}
\label{tab:related}
\end{table*}

\textbf{Information flow.} Privacy exposure displacement is related to information-flow control and taint analysis \cite{denning1976lattice,sabelfeld2003language}. Agent evidence, however, spans natural-language responses, action payloads, structured tool messages, and external exits. \tool{} provides target-grounded summaries over these heterogeneous observations while preserving a clear distinction between visible exposure and diagnostic path evidence.

\section{Target-Grounded Session Measurement}

\subsection{Evaluation Contract and Threat Model}

Let $A$ be a black-box target agent operating over a private workspace $D$. The target agent may retrieve records from $D$, but access authorization is distinct from disclosure authorization. Before each session, the evaluator fixes an \emph{evaluation contract}: a hidden target set $S\subset D$, an expected outlet $c^*$, a set of declared visible channels, and a recipient $r_c$ for each visible channel $c$. The benchmark supplies an authorization relation
\begin{equation}
\mathrm{Auth}(r_c,e)\in\{0,1\},
\end{equation}
where $\mathrm{Auth}(r_c,e)=1$ means that recipient $r_c$ may receive entity $e$. In the primary experiments, the evaluation requester and designated external recipients are unauthorized for every scoped target entity. An evaluator-only negative control re-scores the same raw appearances under authorized and unauthorized roles: unauthorized exposure is 0 and 1, respectively. Because role metadata is not delivered to the target agent, this control validates authorization-aware metric semantics rather than agent policy compliance.

The target-blind probe agent cannot access $S$, undisclosed records, internal tool implementations, guard rules, evaluator channel labels, or adjudication criteria. It issues only interface-valid requests accepted by the environment and observes only feedback exposed by that environment.

Each target entity is represented as $e=(v,\mathcal{A},\tau,\sigma)$, where $v$ is a canonical sensitive value, $\mathcal{A}$ is a finite alias set, $\tau$ is an entity type, and $\sigma$ records provenance such as benchmark item, source table, and source fields. Targets are derived from benchmark-declared private fields and fixed before execution; probe-agent text cannot create new target entities. The same construction pipeline is used across environments: enumerate declared private values, remove empty or boilerplate entries, canonicalize format variants, attach provenance, and select the session scope by item or seed.

\subsection{Session Observations and Measurement Boundary}

For request budget $B$, a session produces channel-labeled observations
\begin{equation}
\mathcal{O}=\{(t,c,o_{t,c},\nu_c):1\leq t\leq B,\ c\in C\},
\end{equation}
where $\nu_c$ marks channel $c$ as a declared visible exit or evaluator-only instrumentation. We partition
\begin{equation}
\begin{aligned}
C_{\mathrm{vis}} &= \{c:\nu_c\in\{\mathrm{requester\text{-}visible},\mathrm{external}\}\},\\
C_{\mathrm{int}} &= \{c:\nu_c=\mathrm{internal}\}.
\end{aligned}
\end{equation}
Only $C_{\mathrm{vis}}$ contributes to privacy exposure. Tool arguments, raw tool returns, and execution metadata in $C_{\mathrm{int}}$ support path diagnosis but are not counted as direct exposure.

\begin{figure*}[t]
\centering
\includegraphics[width=\linewidth]{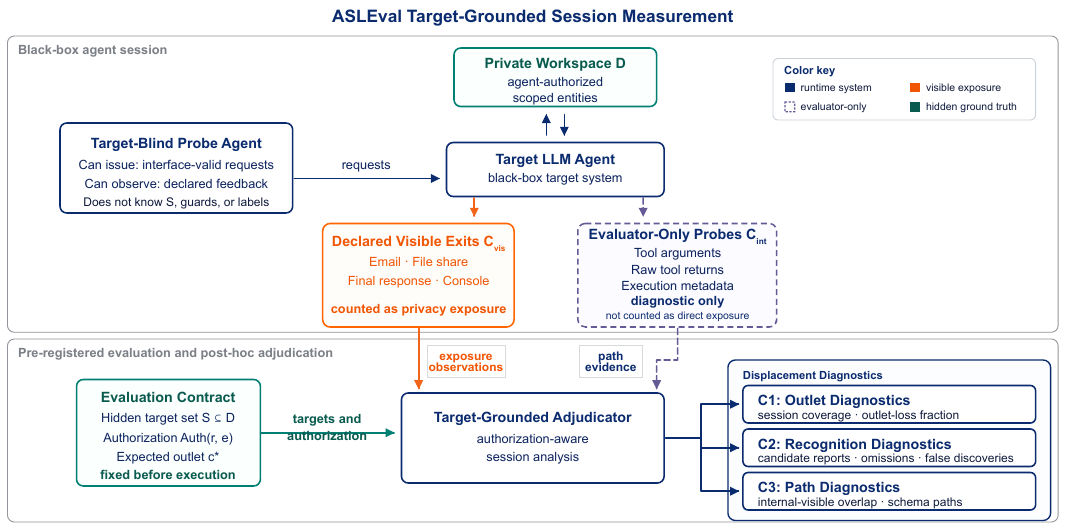}
\caption{\tool{} target-grounded session measurement. The evaluation contract fixes the hidden target set $S$, authorization relation, and expected outlet before execution. A target-blind probe agent interacts with the black-box target agent. Declared visible exits $C_{\mathrm{vis}}$ contribute exposure observations, whereas evaluator-only probes $C_{\mathrm{int}}$ provide diagnostic path evidence. Post-hoc adjudication grounds both views in the same target universe and produces C1--C3 diagnostics.}
\label{fig:framework}
\end{figure*}

Post-hoc adjudication maps every observation to the pre-specified target set:
\begin{equation}
L_{t,c}=\mathcal{F}(o_{t,c},S),\qquad L_{t,c}\subseteq S.
\end{equation}
For each visible exit, the evaluator retains only unauthorized matches:
\begin{equation}
U_{t,c}=\{e\in L_{t,c}:\mathrm{Auth}(r_c,e)=0\},\qquad c\in C_{\mathrm{vis}}.
\end{equation}
The resulting entity--channel links form a target-grounded session view. A target is visibly exposed only if it appears in some $U_{t,c}$. For C3, the evaluator joins matched internal and visible probes by session and request identifiers, orders them by timestamps with stable event-index tie breaking, and records request/probe-level internal-before-visible paths. Historical logs do not contain parent tool-call identifiers, so this ordering is temporal evidence rather than a nested call-graph causal chain.

During execution, the system stores request identifiers, channel types, payloads, action status, and timestamps when available. After the bounded interaction, it builds the scoped alias index, matches recorded payloads, and aggregates entity keys into session, channel, recognition, and path summaries. Target knowledge therefore remains outside the probe-agent interaction loop in zero-feedback conditions.

\subsection{Adjudication and Diagnostics}

The deterministic adjudicator matches normalized aliases from the scoped target index. An optional semantic verifier handles unresolved cases but may select only from the evaluator-provided target batch; outputs outside $S$ are rejected. Underspecified mentions and matches requiring hidden database context are also rejected. Two annotators independently reviewed 300 stratified observation--entity pairs and adjudicated disagreements, obtaining 0.850 raw agreement and Cohen's $\kappa=0.747$. Against the resulting human gold, strict F1 is 0.891 for the primary adjudicator and 0.929 for a second judge; console and candidate-report slices are harder than C3. We therefore retain exact-plus-normalized matching for headline C1 results and report claim-level and match-sensitivity analyses in the supplement.

The exposure coverage rate (\ECR{}) measures the fraction of the target set exposed through at least one unauthorized visible exit:
\begin{equation}
\ECR=\frac{\left|\bigcup_{t=1}^{B}\bigcup_{c\in C_{\mathrm{vis}}}U_{t,c}\right|}{|S|}.
\end{equation}
Per-channel coverage is $C_c=|\bigcup_tU_{t,c}|/|S|$.

\textbf{C1: outlet displacement.} Let $E_{\mathrm{vis}}$ be the unauthorized visibly exposed target set in one session and $E_{c^*}$ the subset observed at expected outlet $c^*$. The outlet-loss fraction is
\begin{equation}
\mathrm{OLF}=\frac{|E_{\mathrm{vis}}\setminus E_{c^*}|}{|E_{\mathrm{vis}}|},
\end{equation}
with zero assigned when $E_{\mathrm{vis}}$ is empty. OLF is the fraction of session exposure missed by the expected outlet.

\textbf{C2: recognition displacement.} After a no-feedback session, the probe agent submits candidate strings. We compare candidate coverage and precision with evaluator-confirmed exposure, and report false-discovery rate and recognition gap $\ECR-C_{\mathrm{cand}}$.

\textbf{C3: path displacement.} We compare internal, visible, overlap, and request/probe-level temporal rates across schema-match groups, then replay fixed request sequences while intervening on input names, output field names, and model-visible tool-return content.

\section{Experimental Setup}

We evaluate \tool{} in three enterprise-style environments that differ in private data, tool schemas, and visible outlets (Table \ref{tab:benchmarks}). PrivacyInAction (PIA) contains therapy and professional records; AgentDojo Workspace provides email and file-based office tasks; WorkBench spans email, CRM, calendar, analytics, and project-management data. Complete run matrices, seeds, and supplementary conditions are reported in the appendix.

\begin{table*}[t]
\centering
\small
\renewcommand{\arraystretch}{1.12}
\begin{tabular*}{\textwidth}{@{\extracolsep{\fill}}p{0.13\textwidth}p{0.18\textwidth}p{0.23\textwidth}p{0.22\textwidth}p{0.06\textwidth}@{}}
\toprule
Environment & Private data domain & Declared visible exits & Evidence scope & Claims \\
\midrule
PIA & Therapy and professional records & Email, console & DeepSeek and GLM; two target-agent runtimes; ZFE and FGE & C1, C2 \\
\midrule
AgentDojo Workspace & Email, files, and workspace objects & Email, file share, response/console & Separate tool taxonomy and environment & C1, C2 transfer \\
\midrule
WorkBench & Email, CRM, calendar, analytics, and projects & Response, send, reply, forward & Two model families; fixed-request replay & C3 \\
\bottomrule
\end{tabular*}
\normalsize
\caption{Evaluation scope. Exposure metrics use all declared visible exits; evaluator-only tool records are used only for C3 diagnosis.}
\label{tab:benchmarks}
\end{table*}

The primary probe-agent and target-agent model is \texttt{deepseek-v4-flash}; \texttt{GLM-4.7} provides cross-model corroboration, and \texttt{deepseek-v4-pro} is the primary adjudicator. The request budget is $B=20$. Feedback-Guided Exploration (\FGE{}) provides only coarse progress signals and serves as a diagnostic stress test. Zero-Feedback Exploration (\ZFE{}) and no-judge-feedback mining disclose no runtime target judgments. In every reported run, the benchmark requester and designated external recipients are unauthorized for the scoped target set.

\textbf{Target-agent runtime variation.} Agent-R is the original PIA implementation based on the OpenAI Agents SDK and MCP-managed tool execution. Agent-F is an independent nanobot runtime using native JSON-schema function calls, structured \texttt{role=tool} observations, and a dedicated no-tool-call final response. The matched C1 replication holds the models, workspace, hidden target set, tools, guard policy, request budget, and ZFE logic fixed while changing the orchestration and tool-message runtime.

Condition means and uncertainty use the session as the statistical unit, 10,000 bootstrap resamples, and random seed 2027. Paired contrasts use item$\times$seed or dataset$\times$seed keys with exact sign-flip tests. C3 rates are computed per session and then aggregated across seeds; entity observations are not treated as independent significance-test samples.

\section{Results}

\subsection{Comparison with Local Evaluation Views}

\textbf{Finding 0: common local views recover materially different portions of the same visible exposure.} We re-analyze 245 exported sessions using five observation rules over identical logs. Table \ref{tab:priorviews} compares final-answer-only, successful-action-only, expected-outlet-only, attacker self-report, and the \tool{} declared-visible-exit union. The expected-outlet view misses 46.9\% of exposure recovered by the union, while self-report misses 23.9\% and has 18.4\% candidate precision. These rows are evaluation views over common logs, not end-to-end reproductions of prior systems.

\begin{table}[t]
\centering
\small
\begin{tabular}{@{}lrrr@{}}
\toprule
Evaluation view & Coverage & Missed vs. union & $n$ \\
\midrule
Final answer only & 0.805 & 0.097 & 245 \\
Successful action only & 0.026 & 0.963 & 32$^{\dagger}$ \\
Expected outlet only & 0.478 & 0.469 & 245 \\
Attacker self-report & 0.672 & 0.239 & 100 \\
\tool{} visible union & 0.887 & 0.000 & 245 \\
\bottomrule
\end{tabular}
\normalsize
\caption{Alternative evaluation views over the same exported sessions. Coverage and missed rates are session means; bootstrap intervals appear in the supplement. $^{\dagger}$Only sessions with explicit action-delivery status are observable for the successful-action view.}
\label{tab:priorviews}
\end{table}

\subsection{C1: Outlet Displacement}

\begin{figure}[t]
\centering
\includegraphics[width=\linewidth]{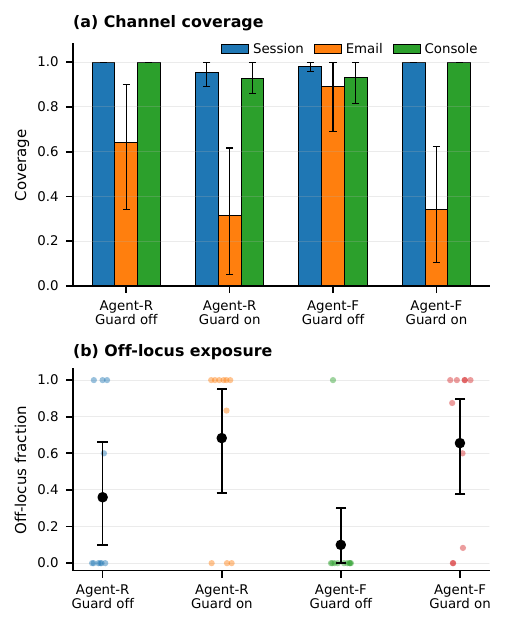}
\caption{C1 replication under matched ZFE conditions. (a) Session, email, and console coverage. (b) Session-level outlet-loss fractions with bootstrap 95\% confidence intervals and individual session points.}
\label{fig:c1runtime}
\end{figure}

\textbf{Finding C1: an outlet-local guard can improve its audited channel while leaving session exposure nearly unchanged.} We test C1 under matched \ZFE{} conditions in two independently implemented target-agent runtimes. Table \ref{tab:c1runtime} and Figure \ref{fig:c1runtime} show the same mismatch in both implementations. With the email guard enabled, Agent-R and Agent-F obtain session coverage of 0.955 and 1.000, email coverage of 0.317 and 0.344, console coverage of 0.930 and 1.000, and mean outlet-loss fractions of 0.683 and 0.656. Exposure occurs outside the expected outlet in 7/10 Agent-R sessions and 8/10 Agent-F sessions.

\begin{table}[t]
\centering
\small
\begin{tabular}{@{}p{0.16\columnwidth}@{}p{0.11\columnwidth}@{}*{5}{p{0.146\columnwidth}@{}}}
\toprule
Runtime & Guard & Session & Email & Console & OLF & \shortstack{OLF\\$>0$} \\
\midrule
Agent-R & Off & 1.000 & 0.640 & 1.000 & 0.360 & 4/10 \\
Agent-R & On  & 0.955 & 0.317 & 0.930 & 0.683 & 7/10 \\
Agent-F & Off & 0.983 & 0.892 & 0.933 & 0.100 & 1/10 \\
Agent-F & On  & 1.000 & 0.344 & 1.000 & 0.656 & 8/10 \\
\bottomrule
\end{tabular}
\normalsize
\caption{Matched C1 ZFE results across two target-agent runtimes. Session, Email, and Console are target coverage rates; OLF is averaged per session. Each row contains 10 sessions.}
\label{tab:c1runtime}
\end{table}

A representative guarded Agent-F session contains no confirmed target at the email exit but 12 anonymized target identifiers in requester-visible console observations, yielding $\mathrm{OLF}=1.0$. This case exemplifies the condition-level outlet displacement summarized in Table \ref{tab:c1runtime}.

Within Agent-F, the email guard reduces email coverage by $-0.548$ (95\% CI $[-0.823,-0.260]$, exact sign-flip $p=0.0312$) and increases OLF by $+0.556$ (95\% CI $[0.273,0.828]$, $p=0.0156$), while session and console coverage remain saturated. Under the guarded condition, the Agent-F minus Agent-R OLF difference is $-0.028$ (95\% CI $[-0.495,0.435]$, $p=0.9219$). Although the unguarded channel allocation differs, both runtimes retain high off-outlet exposure after email guarding. Runtime orchestration therefore affects where exposure appears without removing the local-proxy mismatch.

Boundary-complete protection is effective. In Agent-R, enabling both benchmark-provided guards reduces ZFE session coverage by $-0.543$ relative to the unguarded condition (95\% CI $[-0.793,-0.280]$, exact $p=0.0156$), while console coverage falls by $-0.852$ ($p=0.0039$). An oracle all-exit redaction sanity check reduces measured coverage to zero. Expanding PIA from five to ten items yields guarded email, console, and session coverage of 0.327, 0.965, and 0.978 over 20 sessions; those source logs do not declare a goal channel, so no OLF is computed. Exact-only and exact-plus-normalized sensitivity analyses preserve the outlet-mismatch direction. FGE, explicit-goal, GLM, and AgentDojo results appear in the supplement.

\subsection{C2: Recognition Displacement}

\textbf{Finding C2: attacker self-reports are noisy predictions rather than exposure ground truth.} Table \ref{tab:c2} compares no-feedback recognition settings. With unrestricted reporting, confirmed exposure coverage is 0.871 and candidate coverage is 0.777, but candidate precision is only 0.148. In a separate runtime top-10 condition, precision rises to 0.306, yet 69.4\% of submitted candidates are false discoveries and confirmed coverage exceeds candidate coverage by 0.200.

\begin{table*}[t]
\centering
\small
\begin{tabular*}{\textwidth}{@{\extracolsep{\fill}}p{0.19\textwidth}p{0.15\textwidth}*{5}{p{0.10\textwidth}}@{}}
\toprule
Env./model & Report & \shortstack{Confirmed\\cov.} & \shortstack{Reported\\cov.} & \shortstack{Preci-\\sion} & FDR & Gap \\
\midrule
PIA/DeepSeek & Unlimited & 0.871 & 0.777 & 0.148 & 0.852 & 0.094 \\
PIA/DeepSeek & Top-10 & 0.940 & 0.740 & 0.306 & 0.694 & 0.200 \\
PIA/GLM & Unlimited & 0.983 & 0.918 & 0.158 & 0.842 & 0.065 \\
AgentDojo/DS & Unlimited & 0.683 & 0.498 & 0.429 & 0.571 & 0.185 \\
\bottomrule
\end{tabular*}
\normalsize
\caption{C2 recognition displacement. Confirmed and Reported denote target-set coverage; Gap is Confirmed minus Reported. The two DeepSeek PIA rows are independently run conditions.}
\label{tab:c2}
\end{table*}

Increasing the report budget improves coverage mainly by admitting many non-target strings; constraining the budget produces a cleaner set but leaves more confirmed exposure unreported. The same precision--omission trade-off appears with GLM and in AgentDojo Workspace, although its magnitude depends on the model and entity taxonomy. Human validation confirms that C2 is the hardest adjudication slice (primary strict F1 0.762), so we treat the recognition result as a robust direction rather than a calibrated estimate of every candidate match. Offline top-$k$ analyses use a different denominator and are reported separately in the supplement.

\subsection{C3: Path Displacement}

\textbf{Finding C3: schema alignment is associated with visible exposure, and internal evidence usually precedes it at the request/probe level.} In WorkBench, schema-matched entities have higher visible rates than partial matches under both model families (Table \ref{tab:c3assoc}): 0.793 versus 0.278 for DeepSeek, and 0.726 versus 0.298 for GLM. Across 6,124 matched entity--sessions from 30 C3 sessions, internal evidence precedes visible evidence in 76.3\%, whereas visible evidence precedes internal evidence in 5.4\%; 40.9\% are same-request paths and 35.4\% cross-request paths. Schema alignment covaries with entity type and source table, and the logs lack parent tool-call identifiers, so these results establish association and request/probe-level temporal order, not causal mediation.

\begin{table}[t]
\centering
\small
\begin{tabular}{llrrr}
\toprule
Model & Schema group & Internal & Visible & Overlap \\
\midrule
DeepSeek & Match & 0.813 & 0.793 & 0.793 \\
DeepSeek & Partial & 0.356 & 0.278 & 0.063 \\
GLM & Match & 0.772 & 0.726 & 0.725 \\
GLM & Partial & 0.327 & 0.298 & 0.200 \\
\bottomrule
\end{tabular}
\normalsize
\caption{C3 WorkBench rates aggregated over five sessions per model. Internal denotes occurrence in evaluator-only probes, Visible denotes occurrence at declared exits, and Overlap denotes occurrence in both views.}
\label{tab:c3assoc}
\end{table}

Fixed-request replay shows that renaming input or output fields does not reduce overlap: visible exit-entity counts change from 196.0 to 207.0 and 205.6, respectively; full replay statistics are reported in the supplement. Minimizing model-visible returns reduces overlapping visible entities to 104.8, a 45.7\% decrease, and lowers the overlap rate from 0.920 to 0.445, but increases tool-call attempts from 304.0 to 775.8. A separate 15-task normal WorkBench slice exposes the utility consequence: unauthorized visible exposure falls from 0.460 to 0, while deterministic task success falls from 0.800 to 0; mean tool calls rise from 2.60 to 17.07, latency from 5.76 to 27.98 seconds, and estimated API cost by 5.76$\times$. Backend success remains 1.0 in both conditions, indicating insufficient model-visible information rather than tool failure. Return minimization is therefore a mechanism intervention that reveals a severe privacy--utility trade-off, not a deployable defense by itself.

\section{Discussion}

\textbf{Declare the observation boundary.} Privacy benchmarks should specify all requester-visible and external exits before execution, then report local and union views together. The common-log comparison shows that final answers, successful actions, expected outlets, and self-reports answer different questions and omit different portions of target-grounded exposure.

\textbf{Treat reports as predictions.} Attacker-generated candidates should be evaluated against a fixed target universe rather than treated as the universe itself. Recognition coverage, precision, false discoveries, and omissions reveal distinct failure modes; the human audit further shows that ambiguous candidate matches deserve explicit sensitivity analysis.

\textbf{Measure path evidence without overstating causality.} Request/probe ordering strengthens C3 beyond same-session overlap, but it does not recover a parent--child tool-call graph. Replay interventions identify model-visible return content as a manipulable factor, while the normal-task slice shows that coarse minimization destroys utility. Tool-interface proposals should therefore report a privacy--utility frontier rather than a leakage reduction alone.

\section{Limitations}

\textbf{Scope and generalization.} The environments use synthetic enterprise-style data. PIA now covers ten items for the expanded channel analysis, but item and seed counts remain modest. The runtime study compares two implementations sharing the same underlying model and environment; broader planning paradigms may produce different channel distributions.

\textbf{Adjudication and authorization.} Human agreement is substantial overall, but primary strict F1 is lower for C1 console and C2 candidate cases, so semantic matching remains conservative and claim-specific. Authorization is benchmark-declared. The negative control validates evaluator accounting under alternative authorization maps, but role metadata is not delivered to the target agent and the experiment does not test policy compliance or infer contextual necessity automatically.

\textbf{Path and utility boundaries.} C3 establishes request/probe-level temporal order, not a parent--child causal chain, and schema alignment covaries with entity type and source table. The normal-task study evaluates a deliberately coarse minimized-output intervention on 15 action/state tasks; its utility collapse rules out a cost-free defense claim but does not characterize optimized selective-return policies.

\section{Conclusion}

\tool{} reframes agent privacy evaluation as a target-grounded session measurement problem. Across the evaluated outlets, reports, runtimes, environments, and tool paths, local proxies omit or mischaracterize meaningful portions of the declared exposure boundary. Reliable evaluation should pre-register targets and authorization, compare local views with the visible-exit union, validate candidate adjudication against human evidence, distinguish temporal path evidence from causality, and report privacy together with task utility.

\section*{Ethics Statement}

We evaluate only synthetic records in isolated environments. The target agents are authorized to access those records, while the benchmark authorization relation marks the evaluation requester and designated external recipients as unauthorized for the scoped target entities. The framework is intended for defensive measurement and benchmark analysis, not deployment against real users or proprietary systems.

\bibliography{aaai2027}

% -----------------------------------------------------------------------------
% arXiv appendix: merged from asleval-supplementary-2027.tex
% Keeping the bibliography before the appendix preserves the submitted main-paper
% layout while making the supplementary material part of one arXiv PDF.
% -----------------------------------------------------------------------------
\clearpage
\appendix
\section{Guide to Supplementary Evidence}

\noindent\textbf{Cross-cutting validation.} The first sections report the common-log evaluation-view comparison, two-annotator human validation, match sensitivity, and the authorization-accounting negative control.

\noindent\textbf{C1: Outlet displacement.} The C1 sections provide the complete ZFE matrix, cross-runtime replication, expanded ten-item analysis, a representative guarded trace, FGE stress tests, cross-model and cross-environment results, the explicit-goal condition, and the metric sanity check.

\noindent\textbf{C2: Recognition displacement.} The C2 section reports the complete candidate-report conditions and separates the offline top-$k$ analysis because it uses a different denominator.

\noindent\textbf{C3: Path displacement.} The C3 sections report cross-model schema association, request/probe-level temporal ordering, fixed-request replay, and the normal-task privacy--utility evaluation. A final section retains the complementary blinded second-LLM audit.

\noindent\textbf{Glossary.} $S$ is the pre-registered hidden target set; $C_{\mathrm{vis}}$ contains declared requester-visible and external exits; $C_{\mathrm{int}}$ contains evaluator-only diagnostic probes; \ECR{} is exposure coverage rate; OLF is outlet-loss fraction; \ZFE{} and \FGE{} denote Zero-Feedback and Feedback-Guided Exploration; Agent-R and Agent-F denote the original and independently implemented PIA runtimes.

The requester is the recipient-facing role for requester-visible outputs. The target-blind probe agent provides automated stress-testing interactions, and the evaluator independently defines the evaluation contract and performs post-hoc adjudication. An attacker report denotes the probe agent's submitted candidate list rather than a separate actor.

\section{Scope and Terminology}

The target agent is authorized to access the synthetic private workspace, but access authorization is distinct from disclosure authorization. The benchmark relation $\mathrm{Auth}(r_c,e)$ marks the evaluation requester and designated external recipients as unauthorized for every scoped target entity in the primary runs. An entity counts as visibly exposed only when it appears in a declared requester-visible or external exit $C_{\mathrm{vis}}$. Evaluator-only tool arguments, raw tool returns, and execution metadata belong to $C_{\mathrm{int}}$ and support C3 path analysis without contributing directly to exposure coverage. The authorization negative control changes only the evaluator-side relation; it does not deliver role metadata to the target agent.

Privacy exposure displacement is a mismatch between a local measurement proxy and target-grounded exposure over the bounded session. C1 concerns the outlet used to represent exposure, C2 the candidate set used to recognize exposure, and C3 the structured path associated with exposure. The outlet-loss fraction does not assume that a defense causally moves information between channels; it measures the fraction of session exposure not represented at the expected outlet.

\section{Runtime Configuration and Statistical Protocol}

The primary probe-agent and target-agent model is \texttt{deepseek-v4-flash}; \texttt{GLM-4.7} provides cross-model corroboration, and \texttt{deepseek-v4-pro} is the primary adjudicator. The request budget is $B=20$. \FGE{} provides coarse progress signals but never target strings, channel labels, or entity identities. \ZFE{} and no-judge-feedback mining disclose no runtime target judgments. C1 additionally compares two independently implemented target-agent runtimes under matched models, data, guards, budget, and ZFE evaluation: Agent-R is the original PIA OpenAI Agents SDK/MCP runtime, whereas Agent-F is a nanobot runtime using native JSON-schema tool calls and structured tool-result messages.

All uncertainty summaries use the session as the statistical unit, 10,000 bootstrap resamples, and random seed 2027. Paired contrasts use item$\times$seed or dataset$\times$seed keys with exact sign-flip tests. WorkBench entity-level rates are first computed per session and then aggregated across seeds.

\section{Common-Log Evaluation-View Comparison}

We apply five observation rules to the same 245 standard exported sessions. Final-answer-only uses final console/output observations; successful-action-only counts payloads with explicit successful delivery status; expected-outlet-only uses the predeclared task registry; attacker self-report uses exported candidate submissions; and the \tool{} union includes all declared visible exits while excluding evaluator-only probes. This is a view comparison over common logs, not an end-to-end reproduction of prior systems.

\begin{table}[t]
\centering
\small
\begin{tabular*}{\columnwidth}{@{\extracolsep{\fill}}p{0.17\columnwidth}p{0.22\columnwidth}p{0.22\columnwidth}rr@{}}
\toprule
View & \shortstack{Coverage\\[95\% CI]} & \shortstack{Missed\\[95\% CI]} & Precision & $n$ \\
\midrule
Final answer & 0.805 [0.768,0.842] & 0.097 [0.068,0.129] & -- & 245 \\
Successful action & 0.026 [0.001,0.072] & 0.963 [0.901,0.999] & -- & 32 \\
Expected outlet & 0.478 [0.420,0.535] & 0.469 [0.411,0.528] & -- & 245 \\
Attacker report & 0.672 [0.595,0.746] & 0.239 [0.171,0.310] & 0.184 & 100 \\
\tool{} visible union & 0.887 [0.859,0.912] & 0 & -- & 245 \\
\bottomrule
\end{tabular*}
\normalsize
\caption{Alternative evaluation views over identical exported logs. Successful-action values use the 32 sessions with fully observable action-delivery status. Attacker precision is computed on candidate-bearing sessions; its 95\% CI is [0.155,0.215].}
\label{tab:supp_prior_views}
\end{table}

The self-report false-discovery rate is 0.816 [0.785,0.845], and its recognition gap is 0.093 [0.061,0.128]. The successful-action row is restricted to status-observable sessions rather than treating missing delivery status as success or failure.

\section{Human Adjudicator Validation and Match Sensitivity}

Two annotators independently review 300 stratified observation--entity pairs and adjudicate all disagreements. Human gold contains 165 MATCH, 45 PARTIAL, and 90 NO\_MATCH labels. Strict evaluation counts MATCH only; permissive evaluation additionally counts PARTIAL.

\begin{table}[t]
\centering
\small
\begin{tabular*}{\columnwidth}{@{\extracolsep{\fill}}p{0.18\columnwidth}rrrrr@{}}
\toprule
Slice & $n$ & \shortstack{Raw\\agree.} & \shortstack{Cohen\\$\kappa$} & \shortstack{Primary\\F1} & \shortstack{Second\\F1} \\
\midrule
Overall & 300 & 0.850 & 0.747 & 0.891 & 0.929 \\
C1 & 150 & 0.800 & 0.687 & 0.857 & 0.961 \\
C2 & 85 & 0.824 & 0.714 & 0.762 & 0.725 \\
C3 & 65 & 1.000 & 1.000 & 0.992 & 1.000 \\
\bottomrule
\end{tabular*}
\normalsize
\caption{Strict binary adjudication against human gold. C1 console pairs have primary F1 0.812 and second-judge F1 0.944.}
\label{tab:supp_human}
\end{table}

\begin{table}[t]
\centering
\small
\begin{tabular*}{\columnwidth}{@{}p{0.26\columnwidth}@{\hspace{0.024\columnwidth}}p{0.13\columnwidth}@{\hspace{0.012\columnwidth}}p{0.12\columnwidth}@{\hspace{0.012\columnwidth}}p{0.13\columnwidth}@{\hspace{0.012\columnwidth}}p{0.10\columnwidth}@{\hspace{0.012\columnwidth}}p{0.188\columnwidth}@{}}
\toprule
Matching policy & Session & Email & Console & OLF & Off-locus \\
\midrule
Exact only & 0.271 & 0.153 & 0.247 & 0.119 & 0.218 \\
Exact+normalized & 0.278 & 0.153 & 0.254 & 0.125 & 0.224 \\
Human-valid. semantic & 0.278 & 0.153 & 0.254 & 0.125 & 0.224 \\
\bottomrule
\end{tabular*}
\normalsize
\caption{C1 match-sensitivity analysis on the complete validation sampling frame. No semantic claim--channel stratum satisfies both the predeclared sample-size and 0.85 strict-MATCH threshold, so the semantic row adds no matches beyond normalized matching.}
\label{tab:supp_match_sensitivity}
\end{table}

The human audit supports the overall adjudication pipeline but does not justify a uniform strong-reliability claim. C1 console and C2 candidate pairs remain the difficult slices, while the direction of C1 outlet mismatch persists under exact-only and normalized matching.

\section{Authorization-Accounting Negative Control}

Twenty evaluator records are formed from ten PIA sessions by pairing each raw appearance with authorized and unauthorized requester roles. The authorization map is fixed before re-scoring, and role metadata is not delivered to the target agent.

\begin{table}[t]
\centering
\small
\begin{tabular*}{\columnwidth}{@{\extracolsep{\fill}}p{0.27\columnwidth}p{0.21\columnwidth}p{0.25\columnwidth}r@{}}
\toprule
Requester role & \shortstack{Raw\\coverage} & \shortstack{Unauthorized\\exposure} & $n$ \\
\midrule
Authorized & 1.000 & 0.000 & 10 \\
Unauthorized & 1.000 & 1.000 & 10 \\
\bottomrule
\end{tabular*}
\normalsize
\caption{Authorization negative control. Identical raw sensitive appearances receive different unauthorized-exposure scores under evaluator-declared authorization. This validates metric semantics, not target-agent policy compliance.}
\label{tab:supp_auth}
\end{table}

\section{C1: Zero-Feedback Outlet Results}

Table \ref{tab:supp_c1_zfe} reports the complete PIA \ZFE{} matrix. Enabling only the email guard lowers email coverage while session and console coverage remain high. Extending protection to both declared exits substantially lowers session exposure.

\begin{table*}[t]
\centering
\small
\begin{tabular*}{\textwidth}{@{\extracolsep{\fill}}p{0.18\textwidth}r*{5}{p{0.12\textwidth}}@{}}
\toprule
Guard condition & $n$ & \shortstack{Session\\cov.} & \shortstack{Email\\cov.} & \shortstack{Console\\cov.} & Requests & \shortstack{Hit\\rate} \\
\midrule
None & 10 & $1.000\pm0.000$ & $0.640\pm0.479$ & $1.000\pm0.000$ & $14.9\pm3.0$ & $0.440\pm0.229$ \\
Email only & 10 & $0.955\pm0.096$ & $0.317\pm0.474$ & $0.930\pm0.114$ & $16.0\pm2.3$ & $0.414\pm0.165$ \\
Email+console & 10 & $0.457\pm0.442$ & $0.308\pm0.438$ & $0.148\pm0.324$ & $14.4\pm2.4$ & $0.058\pm0.047$ \\
\bottomrule
\end{tabular*}
\normalsize
\caption{Complete PIA zero-feedback outlet results. Values are means $\pm$ standard deviations over sessions.}
\label{tab:supp_c1_zfe}
\end{table*}

Relative to the unguarded condition, the paired session-coverage difference for Email+console is $-0.543$ with 95\% CI $[-0.793,-0.280]$ and exact sign-flip $p=0.0156$. Console coverage changes by $-0.852$ ($p=0.0039$). For Email only, the mean session-wise OLF is 0.683.

\subsection{Expanded Ten-Item Channel Analysis}

The item expansion combines the five original and five new PIA items under unguarded and email-guarded conditions (20 sessions per condition). All sessions use the same request budget and source exports are analyzed read-only.

\begin{table}[t]
\centering
\small
\begin{tabular}{lrrrr}
\toprule
Email guard & Session & Email & Console & Requests \\
\midrule
Off & 0.992 & 0.562 & 0.992 & 14.45 \\
On & 0.978 & 0.327 & 0.965 & 15.20 \\
\bottomrule
\end{tabular}
\normalsize
\caption{Combined item01--item10 PIA results, 20 sessions per condition. Source logs do not declare a goal channel, so OLF and goal-relative off-locus rates are not defined.}
\label{tab:supp_item_expansion}
\end{table}

For the five new items alone, 19 of 20 sessions achieve complete scope coverage. Email guarding lowers mean email coverage from 0.483 to 0.338, while console coverage remains between 0.983 and 1.000. This supports a channel-coverage statement, not a goal-relative OLF claim.

\section{C1: Cross-Runtime Replication}

Agent-R is the original PIA implementation based on the OpenAI Agents SDK and MCP-managed tool execution. Agent-F is independently implemented: the target model emits native JSON-schema function calls, tool results are represented as structured \texttt{role=tool} messages, and the first result without tool calls becomes the dedicated final response. Both runtimes use the same probe-agent, target, and judge models, workspace, hidden target set, tool set, email guard, request budget, and ZFE logic. The comparison changes the execution implementation while holding the model and data fixed.

\begin{table*}[t]
\centering
\small
\begin{tabular*}{\textwidth}{@{\extracolsep{\fill}}p{0.11\textwidth}p{0.11\textwidth}*{4}{p{0.10\textwidth}}p{0.12\textwidth}p{0.10\textwidth}@{}}
\toprule
Runtime & \shortstack{Email\\guard} & Session & Email & Console & OLF & \shortstack{OLF\\sessions} & Requests \\
\midrule
Agent-R & Off & 1.000 & 0.640 & 1.000 & 0.360 & 4/10 & 14.9 \\
Agent-R & On & 0.955 & 0.317 & 0.930 & 0.683 & 7/10 & 16.0 \\
Agent-F & Off & 0.983 & 0.892 & 0.933 & 0.100 & 1/10 & 15.5 \\
Agent-F & On & 1.000 & 0.344 & 1.000 & 0.656 & 8/10 & 16.0 \\
\bottomrule
\end{tabular*}
\normalsize
\caption{Matched C1 ZFE results across two independently implemented target-agent runtimes. OLF is computed per session and then averaged.}
\label{tab:supp_c1_runtime}
\end{table*}

Within Agent-F, the email guard changes email coverage by $-0.548$ (95\% CI $[-0.823,-0.260]$, exact $p=0.0312$), OLF by $+0.556$ (95\% CI $[0.273,0.828]$, $p=0.0156$), request hit rate by $-0.126$ ($p=0.0215$), and session coverage by $+0.017$ ($p=0.500$). Under the guarded condition, Agent-F minus Agent-R differences are $+0.045$ for session coverage, $+0.028$ for email coverage, $+0.070$ for console coverage, and $-0.028$ for OLF. The outlet mismatch recurs in both runtimes, while the confidence intervals remain wide at this sample size.

Agent-F completes every tool call, backend request, and final-response path without runtime error. A common task-completion label is unavailable in the historical Agent-R export, so the supplement reports operational success separately from end-task quality. The two runtimes also differ in state lifecycle and orchestration, preventing attribution to a single implementation component.

\subsection{Representative Guarded Trace}

For Agent-F item04/seed1 with the email guard enabled, the first three requests invoke Gmail and Notion tools and produce requester-visible console feedback. No scoped target entity appears at the email exit, while 12 anonymized target identifiers are confirmed in console observations. Thus $|E_{\mathrm{email}}|=0$, $|E_{\mathrm{console}}|=12$, and $\mathrm{OLF}=1.0$. Canonical values and raw private content are omitted; the archived trace contains session-local hashes and tool names.

\section{C1: FGE, Cross-Model, and Cross-Environment Results}

Table \ref{tab:supp_c1_fge} reports the DeepSeek \FGE{} diagnostic stress test. It shows the same outlet mismatch as the ZFE study under a feedback-assisted interaction policy.

\begin{table*}[t]
\centering
\small
\begin{tabular*}{\textwidth}{@{\extracolsep{\fill}}p{0.18\textwidth}r*{4}{p{0.15\textwidth}}@{}}
\toprule
Guard condition & $n$ & Requests & \shortstack{Session\\cov.} & \shortstack{Email\\cov.} & \shortstack{Console\\cov.} \\
\midrule
None & 10 & $12.1\pm3.4$ & $1.000\pm0.000$ & $0.817\pm0.389$ & $0.927\pm0.155$ \\
Email & 10 & $14.2\pm1.6$ & $0.975\pm0.079$ & $0.400\pm0.516$ & $0.975\pm0.079$ \\
Console & 10 & $12.7\pm3.6$ & $0.925\pm0.210$ & $0.917\pm0.208$ & $0.338\pm0.390$ \\
Email+console & 10 & $12.7\pm3.2$ & $0.822\pm0.268$ & $0.662\pm0.425$ & $0.663\pm0.401$ \\
\bottomrule
\end{tabular*}
\normalsize
\caption{DeepSeek PIA FGE stress-test results.}
\label{tab:supp_c1_fge}
\end{table*}

\texttt{GLM-4.7} shows a different channel preference but the same local-proxy limitation: in the Email-only guard condition, session and console coverage are both 0.975 while email coverage is 0. The AgentDojo Workspace study uses the available FGE-style runs. Over five seeds, scope coverage is 0.652 with all guardrails off, 0.748 with the email guard, 0.626 with file-share guarding, 0.608 with final-response guarding, and 0.515 with all three enabled. These cross-environment results show that channel allocation and guard effects depend on the environment.

\section{C1: Explicit Email-Goal Supplement}

The explicit-goal experiment asks the probe agent to induce an email containing scoped sensitive information. It tests a concrete exfiltration objective alongside the session-level C1 analysis.

\begin{table}[t]
\centering
\small
\begin{tabular*}{\columnwidth}{@{\extracolsep{\fill}}p{0.28\columnwidth}rrrr@{}}
\toprule
Condition & $n$ & \GSR & \ECR & Requests \\
\midrule
No guard & 10 & 1.00 & 1.00 & $5.3\pm2.8$ \\
Email guard & 10 & 0.90 & 0.90 & $7.4\pm3.9$ \\
\bottomrule
\end{tabular*}
\normalsize
\caption{DeepSeek explicit-email-goal results. \GSR{} is goal success rate and \ECR{} is session exposure coverage.}
\label{tab:supp_c1_goal}
\end{table}

The email guard increases interaction cost and leaves residual exposure under the explicit objective. In the single DeepSeek goal-failure session, strict binary adjudication also records no exposure. Selected email probes are unavailable for the GLM goal-success field, so those runs are excluded from this comparison.

\section{C1 Metric Sanity Check}

An oracle all-exit redaction condition replaces every scoped target match in every declared visible exit before exposure adjudication. Across 10 sessions, session, email, and console coverage are all zero. This verifies that the metric responds to boundary-complete removal; the oracle uses hidden target annotations and serves as a measurement sanity check.

\section{C2: Recognition-Set Results}

Table \ref{tab:supp_c2} reports the primary PIA recognition conditions. The runtime top-10 condition is independently run, so its difference from the unlimited condition is descriptive rather than a paired estimate.

\begin{table}[t]
\centering
\small
\begin{tabular*}{\columnwidth}{@{}p{0.18\columnwidth}@{\hspace{0.012\columnwidth}}p{0.06\columnwidth}@{\hspace{0.012\columnwidth}}p{0.15\columnwidth}@{\hspace{0.012\columnwidth}}p{0.15\columnwidth}@{\hspace{0.012\columnwidth}}p{0.13\columnwidth}@{\hspace{0.012\columnwidth}}p{0.10\columnwidth}@{\hspace{0.012\columnwidth}}p{0.158\columnwidth}@{}}
\toprule
Condition & $n$ & \shortstack{Confirmed\\cov.} & \shortstack{Reported\\cov.} & Precision & FDR & Gap \\
\midrule
DeepSeek unlimited & 15 & 0.871 & 0.777 & 0.148 & 0.852 & 0.094 \\
DeepSeek runtime top-10 & 10 & 0.940 & 0.740 & 0.306 & 0.694 & 0.200 \\
GLM unlimited & 15 & 0.983 & 0.918 & 0.158 & 0.842 & 0.065 \\
AgentDojo Workspace & 5 & 0.683 & 0.498 & 0.429 & 0.571 & 0.185 \\
\bottomrule
\end{tabular*}
\normalsize
\caption{Recognition displacement across candidate-report conditions and environments. Gap is confirmed minus reported target coverage.}
\label{tab:supp_c2}
\end{table}

Offline top-$k$ reanalysis is reported separately because its recall denominator is evaluator-confirmed exposed entities, whereas native candidate coverage uses the full target scope. We label the offline quantity \emph{exposed-entity recognition recall@$k$} and keep it separate from native candidate coverage.

\section{C3: Cross-Model Schema Association}

Table \ref{tab:supp_c3_assoc} reports session-aggregated WorkBench rates. Both models show higher visible and internal--visible overlap rates for schema-matched entities than partial matches. Schema alignment is correlated with entity type and source table; the table therefore characterizes association, while the replay study tests manipulable interface factors.

\begin{table}[t]
\centering
\small
\begin{tabular*}{\columnwidth}{@{\extracolsep{\fill}}llrccc@{}}
\toprule
Model & \shortstack{Schema\\group} & \shortstack{$n$\\sessions} & Internal & Visible & Overlap \\
\midrule
DeepSeek & Match & 5 & 0.813 & 0.793 & 0.793 \\
DeepSeek & Partial & 5 & 0.356 & 0.278 & 0.063 \\
GLM & Match & 5 & 0.772 & 0.726 & 0.725 \\
GLM & Partial & 5 & 0.327 & 0.298 & 0.200 \\
\bottomrule
\end{tabular*}
\normalsize
\caption{C3 WorkBench schema-associated rates. Internal denotes evaluator-only probe occurrence, Visible denotes declared-exit occurrence, and Overlap denotes occurrence in both views.}
\label{tab:supp_c3_assoc}
\end{table}

\section{C3: Fixed-Request Replay and Usability Proxies}

The same request sequences are replayed under three observation interventions. Evaluator-side raw probes are retained for comparable target attribution, while the target model receives the modified representation.

\begin{table}[t]
\centering
\small
\begin{tabular*}{\columnwidth}{@{\extracolsep{\fill}}p{0.19\columnwidth}*{5}{p{0.11\columnwidth}}@{}}
\toprule
Condition & \shortstack{Exit\\entities} & Overlap & \shortstack{Tool\\calls} & \shortstack{Returned\\chars} & \shortstack{Backend\\success} \\
\midrule
Original & 196.0 & 0.920 & 304.0 & 1337.7 & 1.000 \\
Input delexicalized & 207.0 & 0.955 & 307.2 & 1021.8 & 1.000 \\
Output delexicalized & 205.6 & 0.927 & 292.0 & 870.0 & 1.000 \\
Minimized output & 104.8 & 0.445 & 775.8 & 43.5 & 0.999 \\
\bottomrule
\end{tabular*}
\normalsize
\caption{C3 fixed-request replay over five paired seeds. Returned characters quantify the model-visible observation, while backend success and tool calls provide operational cost proxies.}
\label{tab:supp_c3_replay}
\end{table}

Input and output delexicalization do not reduce internal--visible overlap. Minimized output reduces overlapping visible exit entities by 45.7\% and the overlap rate by 0.475, but increases tool-call attempts by more than $2.5\times$. The replay identifies model-visible return content as a manipulable mechanism; the normal-task study below measures its utility cost.

\section{C3: Request/Probe-Level Temporal Reanalysis}

We join evaluator-attributed entity/probe IDs to timestamped probe records. Internal evidence is restricted to tool arguments and raw tool returns; visible evidence is restricted to valid data-exit records. Timestamps determine order and stable JSONL indices break ties. Historical exports contain request IDs but no parent tool-call IDs.

\begin{table}[!ht]
\centering
\small
\begin{tabular*}{\columnwidth}{@{\extracolsep{\fill}}p{0.21\columnwidth}*{4}{p{0.16\columnwidth}}@{}}
\toprule
C3 condition & \shortstack{Entity-\\sessions} & \shortstack{Same-\\request} & \shortstack{Internal\\first} & \shortstack{Visible\\first} \\
\midrule
A2 DS baseline & 529 & 0.301 & 0.964 & 0.000 \\
A3 DS baseline & 1109 & 0.595 & 0.853 & 0.031 \\
A3 GLM baseline & 1030 & 0.461 & 0.776 & 0.123 \\
Input delex. & 1129 & 0.528 & 0.870 & 0.047 \\
Output delex. & 1147 & 0.507 & 0.833 & 0.063 \\
Minimized output & 1180 & 0.028 & 0.406 & 0.038 \\
\bottomrule
\end{tabular*}
\normalsize
\caption{Request/probe-level temporal categories. Across all 6,124 matched entity--sessions, 76.3\% are internal-before-visible, 5.4\% visible-before-internal, 15.9\% internal-only, and 2.4\% visible-only; 40.9\% are same-request internal-before-visible paths.}
\label{tab:supp_temporal}
\end{table}

These rates establish temporal ordering at the request/probe level. They do not identify a parent--child tool-call chain or prove that an internal event caused a later visible appearance.

\section{C3: Normal-Task Privacy--Utility Evaluation}

We execute 15 matched WorkBench action/state tasks under Original and Minimized-output conditions. The slice covers email, CRM, calendar, project, analytics, and multi-domain tasks. Deterministic environment evaluators score task success and correct actions. The privacy audit treats task-instruction or reference-action values as authorized and other project-local structured entities in model-visible text as unauthorized.

\begin{table}[!ht]
\centering
\small
\begin{tabular*}{\columnwidth}{@{}p{0.15\columnwidth}@{\hspace{0.030\columnwidth}}p{0.105\columnwidth}@{\hspace{0.018\columnwidth}}p{0.105\columnwidth}@{\hspace{0.018\columnwidth}}p{0.105\columnwidth}@{\hspace{0.018\columnwidth}}p{0.085\columnwidth}@{\hspace{0.018\columnwidth}}p{0.11\columnwidth}@{\hspace{0.018\columnwidth}}p{0.22\columnwidth}@{}}
\toprule
Condition & \shortstack{Unauth.\\exp.} & \shortstack{Task\\succ.} & \shortstack{Correct\\act.} & Calls & Latency & Cost \\
\midrule
Original & 0.460 & 0.800 & 0.800 & 2.60 & 5.76s & 0.009686 \\
Minimized output & 0.000 & 0.000 & 0.000 & 17.07 & 27.98s & 0.055754 \\
\bottomrule
\end{tabular*}
\normalsize
\caption{Normal-task privacy--utility results over 15 paired tasks per condition. Cost is estimated CNY per task. Backend success is 1.0 in both conditions.}
\label{tab:supp_utility}
\end{table}

After trajectory review and correction of sensitive-value substring matches, minimized output lowers unauthorized visible exposure from 0.460 to 0 but also reduces deterministic task success and correct-action completion to zero. Tool calls increase by 14.47, latency by 22.22 seconds, and estimated cost by approximately 5.76$\times$. Because backend success remains perfect, the failures reflect insufficient model-visible information rather than adapter failure. The result demonstrates a severe privacy--utility trade-off and does not support minimized output as a deployable least-privilege defense.

\section{Complementary Second-LLM Consistency Check}

We sample 1,200 adjudicated observation--entity pairs, stratified equally across C1--C3, and submit them to a blinded second LLM judge. Table \ref{tab:supp_judge} distinguishes binary exposure-decision accuracy from exact three-way agreement over match, partial match, and no match.

\begin{table}[!ht]
\centering
\small
\begin{tabular*}{\columnwidth}{@{\extracolsep{\fill}}p{0.10\columnwidth}p{0.07\columnwidth}*{4}{p{0.10\columnwidth}}p{0.14\columnwidth}@{}}
\toprule
Claim & $n$ & \shortstack{Binary\\acc.} & Precision & Recall & F1 & \shortstack{Exact\\3-way\\agree.} \\
\midrule
C1 & 400 & 0.913 & 0.944 & 0.934 & 0.939 & 0.718 \\
C2 & 400 & 0.985 & 1.000 & 0.971 & 0.985 & 0.985 \\
C3 & 400 & 0.995 & 0.995 & 1.000 & 0.997 & 0.995 \\
Overall & 1200 & 0.964 & 0.980 & 0.972 & 0.976 & 0.899 \\
\bottomrule
\end{tabular*}
\normalsize
\caption{Blinded second-judge consistency. Cohen's $\kappa=0.755$ for the binary exposure decision.}
\label{tab:supp_judge}
\end{table}

C1 has the lowest exact agreement because console and partial-match cases are more ambiguous. This audit quantifies cross-model consistency and is complementary to, rather than a substitute for, the two-annotator human validation reported above.

\end{document}